\documentclass[numreferences]{kluwer}
\pdfoutput=1
\usepackage{graphicx}

\def\col#1#2{\left[\matrix{{#1} \cr {#2}}\right]}
\def\pol{\epsilon}
\def\abs#1{\left|{#1}\right|}
\def\bkt#1{\left({#1}\right)}
\def\od#1{\!\bkt{#1}}
\def\bref#1{(\ref{#1})}
\def\lid#1#2#3{\int_{#2}^{#3}\!d{#1}\,}
\def\real#1{\hbox{Re}\!\left\{#1\right\}}

\begin{document}
\begin{article}
\begin{opening}
\title{Spatial and frequency domain effects of defects in 1D photonic crystal}
\author{Adam~\surname{Rudzi\'nski}$^1$\email{a.rudzinski@elka.pw.edu.pl}}
\author{Anna~\surname{Tyszka-Zawadzka}$^1$}
\author{Pawe\l{}~\surname{Szczepa\'nski}$^{1,2}$}
\institute{$^1$Institute of Microelectronics and Optoelectronics, Warsaw University of~Technology, ul.~Koszykowa~75, 00-662 Warszawa, Poland\\
$^2$National Institute of Telecommunications, ul.~Szachowa~1, 00-894 Warszawa, Poland}
\runningauthor{A. Rudzi\'nski, A. Tyszka-Zawadzka and P. Szczepa\'nski}
\runningtitle{Spatial and frequency domain effects of defects in 1D photonic crystal}
\begin{abstract}
The aim of this paper is to present the analysis of influence of defects in 1D photonic
crystal (PC) on the density of states and simultaneously spontaneous emission,
in both spatial and frequency domains.
In our investigations we use an analytic model of 1D PC with defects.
Our analysis reveals how presence of a defect causes a defect mode to appear.
We show that a defect in 1D PC has local character, being negligible in regions
of PC situated far from the defected elementary cell.
We also analyze the effect of multiple defects, which lead to photonic band gap splitting.
\end{abstract}
\keywords{1D photonic crystal, mode spectrum, defect}
\end{opening}

\section{Introduction}
Photonic crystals (PhCs), since first proposed by Yablonovitch
(\cite{Yabl}) and John (\cite{John}), have attracted considerable
attention. These artificial structures are materials patterned
with a periodicity in dielectric constant, which can create a
range of frequencies (called a photonic band gap) for which light
is forbidden to exist within the interior of the crystal. These
photonic band gaps (PBGs) lend themselves to numerous applications
in linear, nonlinear and quantum optics. It has been predicted and
confirmed experimentally that photonic crystals allow to modify
and control spontaneous emission of an exited atom due to
modification of the density of quantum states (\cite{Zhu,Yang}).
In particular, it is well known, that suppression of spontaneous
emission is important in keeping a system in exited states and can
result in the reduction of noise in optoelectronic devices.
Localized electromagnetic modes can be achieved by introduction of
defects within the photonic crystal lattice (\cite{Kawa,Lee}). The
nature of the defect decides on the shapes and properties of
localized photonic states in the gap. A point defect could act
like a~microcavity, a line defect like a waveguide.

In this paper, we study the effect of defects on the density of states and simultaneously
spontaneous emission rate in one-dimensional photonic crystal. We begin by presenting
an analytical model of density of states which is called the effective resonator model
(Section \ref{Secmodel}). In general it is well known, that the microcavity affects density of states and spontaneous emission rate. In our approach effective cavity model is used to study density of states and spontaneous emission rate in finite photonic crystal. Moreover, the influence of local defects on density of states in spectral and spatial domain is considered. Presented model also takes into consideration direction of propagation
and therefore does not neglect anisotropic nature of photonic crystals.
In Section~\ref{Secms} we briefly derive the analytical expression for mode spectrum which
is related to the density of states. This relation allow to analyze properties of
one-dimensional photonic crystals with defects. The influence of defects in
1D photonic crystal on the density of states and spontaneous emission,
in spatial and frequency domain is investigated in Section~\ref{Secdef}. Conclusions can be
found in Section~\ref{Secconc}.

\section{Effective resonator model}\label{Secmodel}
\begin{figure}
\begin{center}
\includegraphics[width=.7\textwidth]{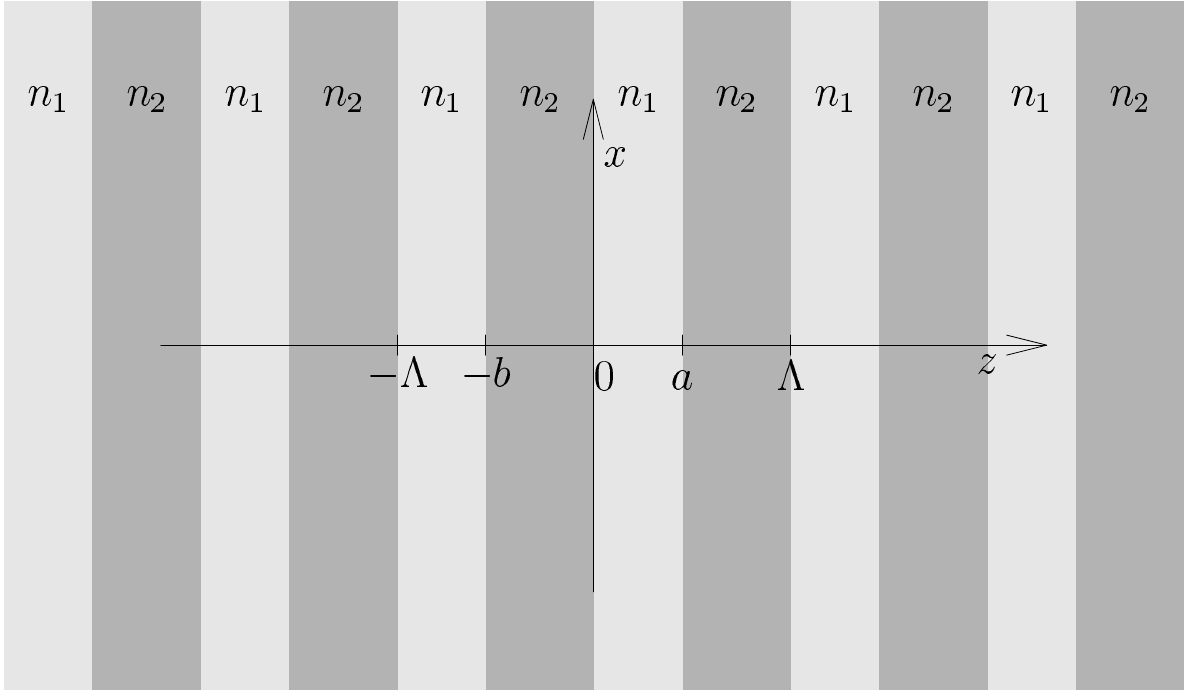}
\caption{Structure of a 1D photonic crystal.}\label{pcstr}
\end{center}
\end{figure}
It is an easy task to calculate reflection coefficient of a
lattice built of dielectric layers, which extend towards infinity
in $x$ and $y$ directions (where coordinate system is taken as
shown in Fig. \ref{pcstr}). Following \cite{Yeh}, in every layer,
for each of two polarizations, a solution of Maxwell equations can
be written as a superposition of two plane waves:
\begin{equation}E_i\od{\vec r}=a_i e^{i\vec k\vec r}+b_i e^{i\vec k_*\vec r}\ ,\end{equation}
where $i$ stands for the number of the layer and
\begin{equation}\vec k_*=\vec k-2k_z\vec e_z\ .\end{equation}
These plane waves are bound with plane waves in the next layers by
continuity conditions, which can be written as a matrix equation (\cite{Yeh, Jack, pinz}):
\begin{equation}\col{a_{i+1}}{b_{i+1}}=M_i\col{a_i}{b_i}\label{conrel},\end{equation}
where $M_i$ is the continuity condition matrix:
\begin{subequation}
\begin{equation}
M_i^{(TE)}=\frac{1}{2}\left[
\matrix{
\bkt{1+\frac{k_z^{(i)}}{k_z^{(i+1)}}}e^{ik_z^{(i)}L_z^{(i)}} &
\bkt{1-\frac{k_z^{(i)}}{k_z^{(i+1)}}}e^{-ik_z^{(i)}L_z^{(i)}}
\cr
\bkt{1-\frac{k_z^{(i)}}{k_z^{(i+1)}}}e^{ik_z^{(i)}L_z^{(i)}} & \bkt{1+\frac{k_z^{(i)}}{k_z^{(i+1)}}}e^{-ik_z^{(i)}L_z^{(i)}}
}
\right]\ ,
\end{equation}
\begin{equation}
M_i^{(TM)}=\frac{1}{2}\left[
\matrix{
\bkt{1+\frac{n_{(i+1)}^2 k_z^{(i)}}{n_{(i)}^2 k_z^{(i+1)}}}e^{ik_z^{(i)}L_z^{(i)}} &
\bkt{1-\frac{n_{(i+1)}^2 k_z^{(i)}}{n_{(i)}^2 k_z^{(i+1)}}}e^{-ik_z^{(i)}L_z^{(i)}}
\cr
\bkt{1-\frac{n_{(i+1)}^2 k_z^{(i)}}{n_{(i)}^2 k_z^{(i+1)}}}e^{ik_z^{(i)}L_z^{(i)}} &
\bkt{1+\frac{n_{(i+1)}^2 k_z^{(i)}}{n_{(i)}^2 k_z^{(i+1)}}}e^{-ik_z^{(i)}L_z^{(i)}}
} \right]
\ ,
\end{equation}
\end{subequation}
where $k_z^{(i)}$ is the wave vector's $z$ component in the $i$th
layer, $L_z^{(i)}$ is the $i$th layer's width and $n_{(i)}$ is the
$i$th layer's refractive index (more precisely, Eq. \bref{conrel}
is a relation between amplitudes of fields in different points of
space -- left edges of layers). Amplitudes of plane waves outside
the lattice are therefore related to each other:
\begin{equation}\col{a_{N+1}}{b_{N+1}}=M_N\ldots M_1M_0\col{a_0}{b_0}\ .\end{equation}
If $a_0$ is an amplitude of the plane wave incident on the
lattice, then $b_{N+1}$ corresponds to the plane wave reflected in
the infinity. This reflection does not occur, hence $b_{N+1}=0$.
Therefore:
\begin{equation}\col{a_{N+1}}{0}=M\col{a_0}{b_0}\ ,\end{equation}
where
\begin{equation}M=\left[\matrix{M_{11} & M_{12} \cr M_{21} & M_{22}}\right]\equiv M_N\ldots M_1M_0\end{equation}
and as a result:
\begin{equation}b_0=-\frac{M_{21}}{M_{22}}a_0\ .\end{equation}
The reflection coefficient of the whole lattice is then given by
\begin{equation}r=\frac{b_0}{a_0}=-\frac{M_{21}}{M_{22}}\ ,\label{refl}\end{equation}
with the assumption, that amplitudes $a_0$ and $b_0$ describe
incident and reflected plane waves in a ``virtual'' layer with the
same refraction coefficient as the external medium and width
$L_z^{(0)}=0$. The above expression is purely analytical, though
for lattices with diversified layers it would take a very
complicated form if written as an explicit function of wave
vector, polarization and lattice parameters, which are hidden in
$M_{21}$ and $M_{22}$.

\begin{figure}
\begin{center}
\includegraphics[width=.7\textwidth]{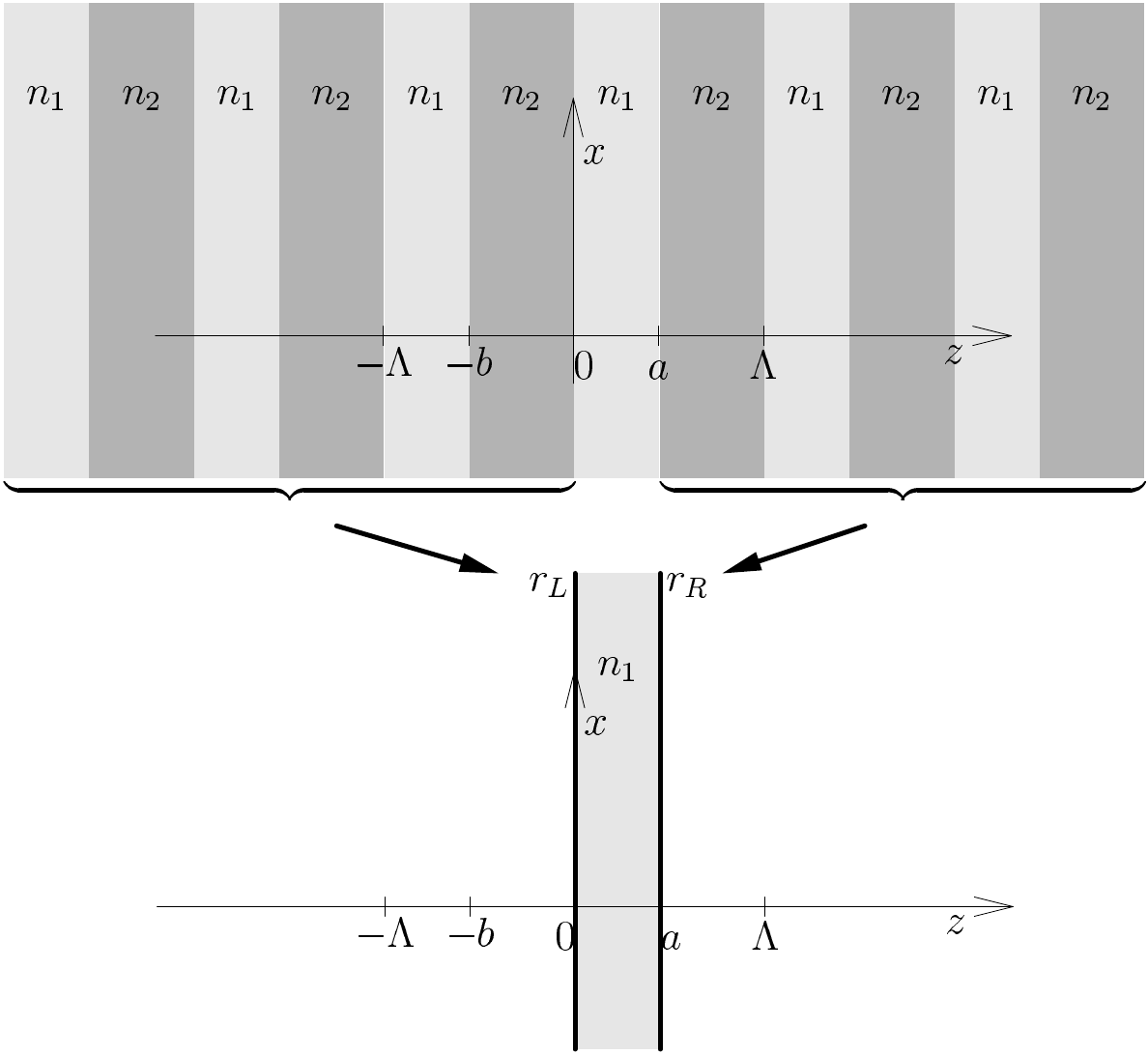}
\caption{Illustration of effective resonator model.}\label{pctor}
\end{center}
\end{figure}
Having derived \bref{refl}, we can calculate reflection
coefficient of any number of dielectric layers, what allows us to
treat any randomly selected layer of a lattice as an effective
resonator. The idea of this model is presented in
Fig.~\ref{pctor}. Formally, a lattice can be treated as two
lattices and the layer joined together. It is possible to replace
the lattices on both sides of the layer with mirrors having the
same reflection coefficients $r_{L\pol}\od{\vec k}$ and
$r_{R\pol}\od{\vec k}$, where $\pol$ denotes polarization.

\section{Mode spectrum}\label{Secms}
Let $\vec E^{ex}$ be a plane wave exciting a resonator limited by two parallel planes
with reflection coefficients $r_L$ and $r_R$ (direction of this vector determines
polarization of the wave). Because of reflections, there appears in
the resonator an effective plane wave $\vec E^{eff}$. We introduce an operator $\hat Q$,
which establishes a relation between them (\cite{pmgr,RudzSPIE}):
\begin{equation}\vec E^{eff}=\hat Q\vec E^{ex}\ .\end{equation}
We call a plane wave $\vec E_\pol\od{\vec k}$ with polarization $\pol$ and wave vector
$\vec k$ a~mode of the resonator, if it is an eigenvector of $\hat Q$:
\begin{equation}\hat Q\vec E_\pol\od{\vec k}=\Lambda_\pol\od{\vec k}\vec E_\pol\od{\vec k}\ .\end{equation}
Eigenvalue $\Lambda_\pol\od{\vec k}$ is proportional to mode spectrum $\rho_\pol\od{\vec k}$,
which we define as:
\begin{equation}\rho_\pol\od{\vec k}=\frac{\Lambda_\pol\od{\vec k}}{8\pi^3}\ .\end{equation}
Mode spectrum describes distribution of modes in wave vector
domain, which is equivalent to frequency and direction of
propagation. We consider the following relation between mode
spectrum and density of states (DOS):
\begin{equation}dN\od{k}=\sum_\pol k^2\bkt{\lid{\Omega}{4\pi}{}\rho_\pol\od{\vec k}}dk\ .\label{DOS}\end{equation}
In free space no reflections occur and therefore the operator
$\hat Q^{fs}$ is an identity, with all eigenvalues equal to~$1$.
Hence, mode spectrum is constant in free space, and with taken
normalization we obtain from \bref{DOS} valid and well known
formula for DOS in free space:
\begin{equation}dN^{fs}\od{k}=\frac{k^2}{\pi^2}dk\ .\end{equation}

Let $\vec E_\pol\od{\vec k}$ be a mode of the considered resonator. If the amplitude
of the excitation is $E_0$, to find the amplitude $E$ of the effective plane wave one has
to sum all the reflected waves:
\begin{equation}E=E_0+\sum_{j=1}^\infty E_0\bkt{r_Lr_R e^{2ik_zL_z}}^j+\sum_{j=1}^\infty E_0\bkt{r_L^*r_R^* e^{-2ik_zL_z}}^j\ ,\end{equation}
where $L_z$ is the resonator's width. The second sum arises
because continuity conditions apply on both mirrors, and it can be
easily checked, that for a plane wave incident on a plane between
two different media the change of sign of $k_z$ (assuming that $z$
axis is perpendicular to the plane) causes the reflection
coefficient to become its conjugate. It follows, that mode
spectrum for the resonator is described with the formula:
\begin{equation}\rho_\pol\od{\vec k}=\frac{1}{8\pi^3}\frac{1-\abs{r_{L\pol}r_{R\pol}}^2}{1+\abs{r_{L\pol}r_{R\pol}}^2-2\real{r_{L\pol}r_{R\pol}e^{2ik_zL_z}}}\ ,\label{modespectrum}\end{equation}
where reflection coefficients depend on polarization and wave vector.

\section{Defects in 1D photonic crystal}\label{Secdef}
\begin{figure}
\begin{center}
\includegraphics[width=.95\textwidth]{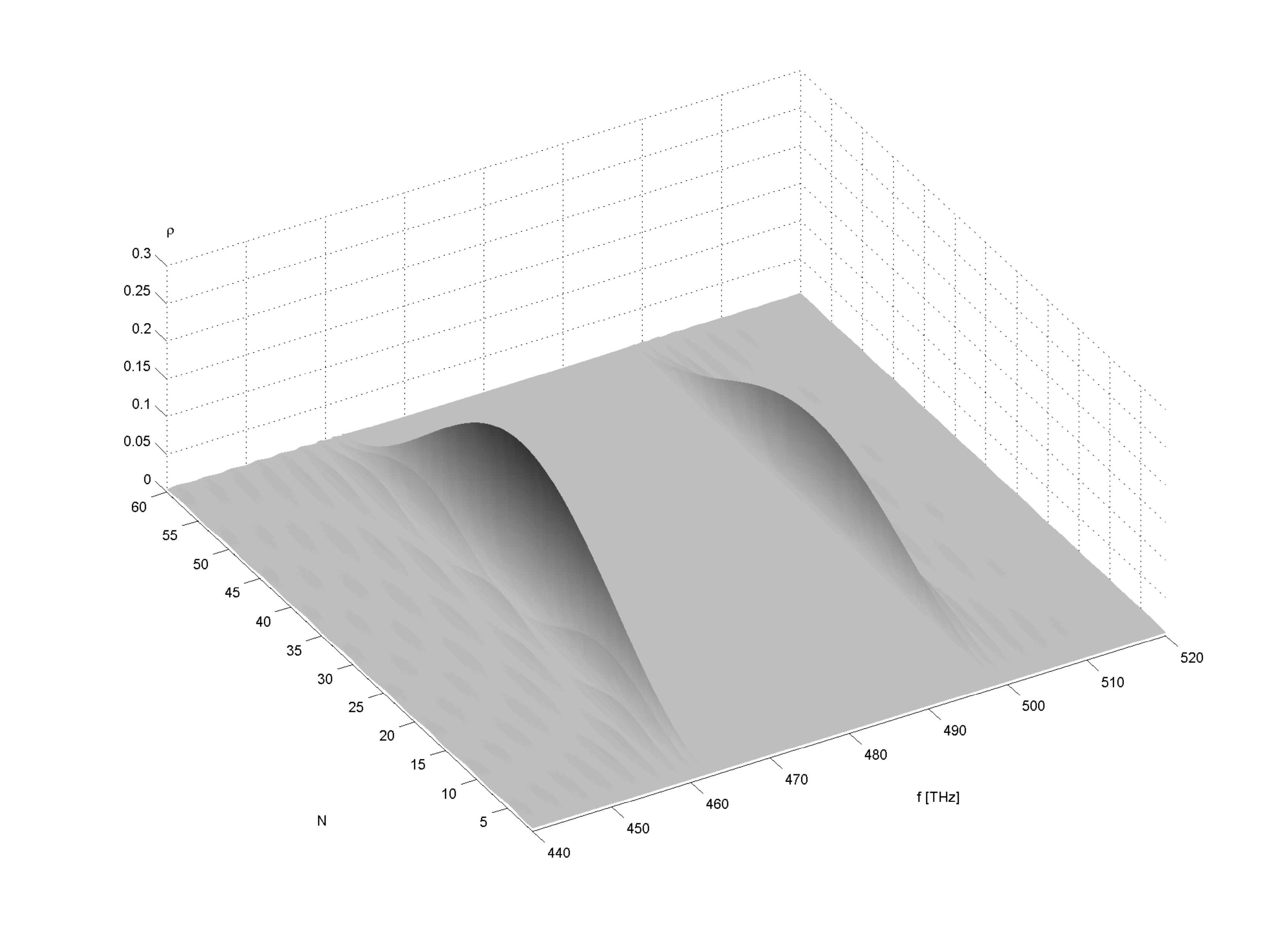}
\caption{Mode spectrum for perpendicular incidence in a PC with no defects versus frequency $f$ and number of elementary cell $N$.}\label{fdef0}
\end{center}
\end{figure}
Formula \bref{modespectrum} allows to analyze properties of layers
of a lattice, in a~particular case -- a 1D photonic crystal. In
this kind of structure there appear photonic band gaps, regions of
frequency for which no mode is allowed. In terms of mode spectrum
a photonic band gap is a~region, for which mode spectrum is equal
to $0$. However, in a structure built of finite number of layers
it is not strictly possible, because the structure is not able to
fully eliminate coupling with electromagnetic field outside of it.
Mode spectrum of a perfect photonic crystal in the region of a
band gap for perpendicular incidence ($\vec k=k_z\vec e_z$) is
shown in Fig. \ref{fdef0} versus frequency $f$ and the number of
elementary cell $N$ in which there is the $n_1$ layer, for which
the calculation has been made. For simplicity, it has been assumed
that the crystal is surrounded by a~uniform medium with refractive
index $n_1$.

As it can be seen in Fig. \ref{fdef0}, for elementary cells near
the middle of the crystal mode spectrum is practically equal to
$0$ in the gap with maximal values at both edges of the gap.
Outside the gap mode spectrum fluctuates around the free space
value. It means, that the layers are resonators with resonant
frequencies at the gap edges. For layers near the crystals edge
there is only a slight difference between mode spectrum of the
layer and a uniform medium (which is constant in the latter case),
because they are resonators with a very low quality and do not
impose a significant change of an amplitude of a plane wave.

\begin{figure}
\begin{center}
\includegraphics[width=.95\textwidth]{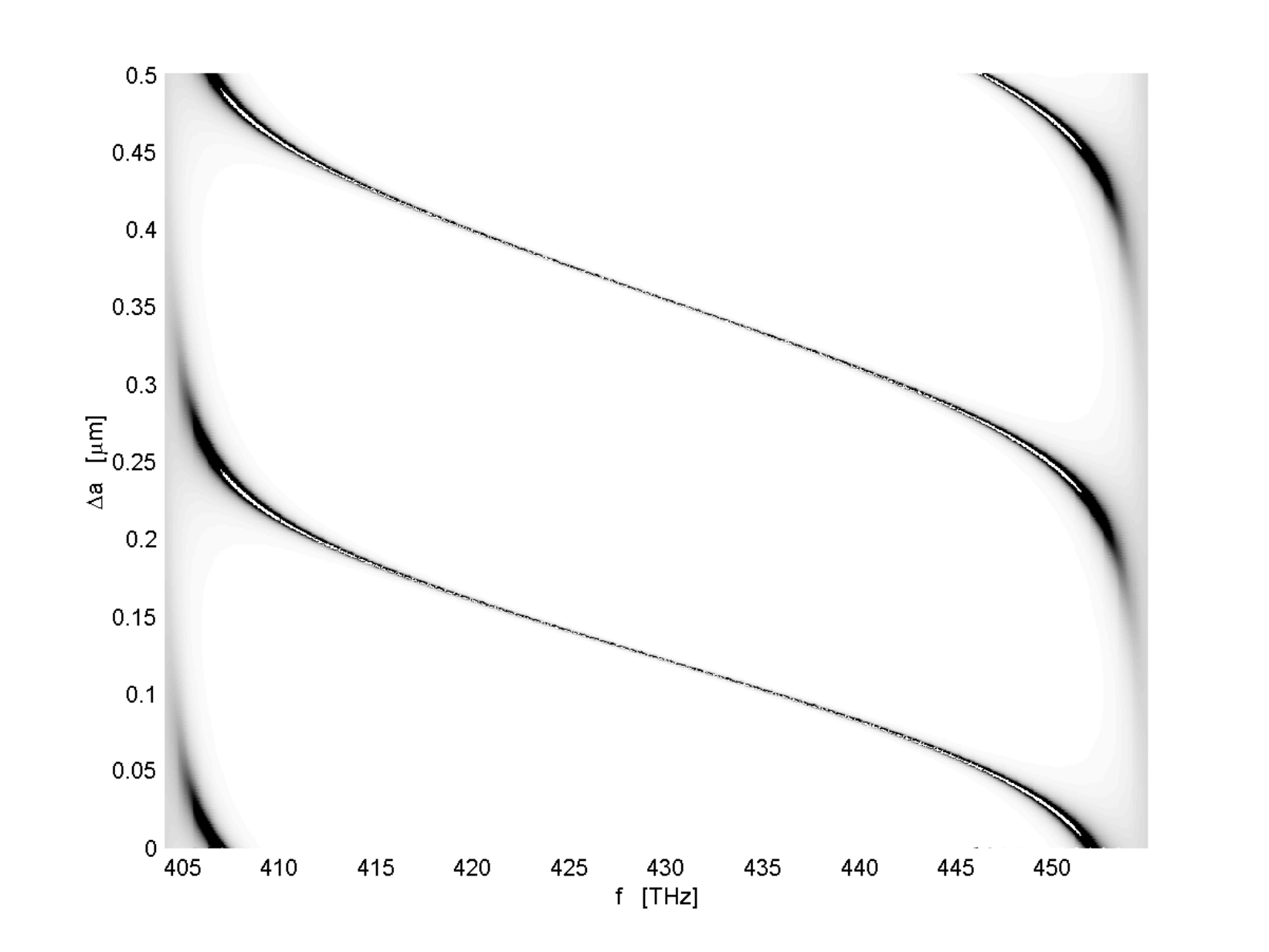}
\caption{Relation between defect mode frequency and defect size.}\label{deffreq}
\end{center}
\end{figure}
\begin{figure}
\begin{center}
\includegraphics[width=.95\textwidth]{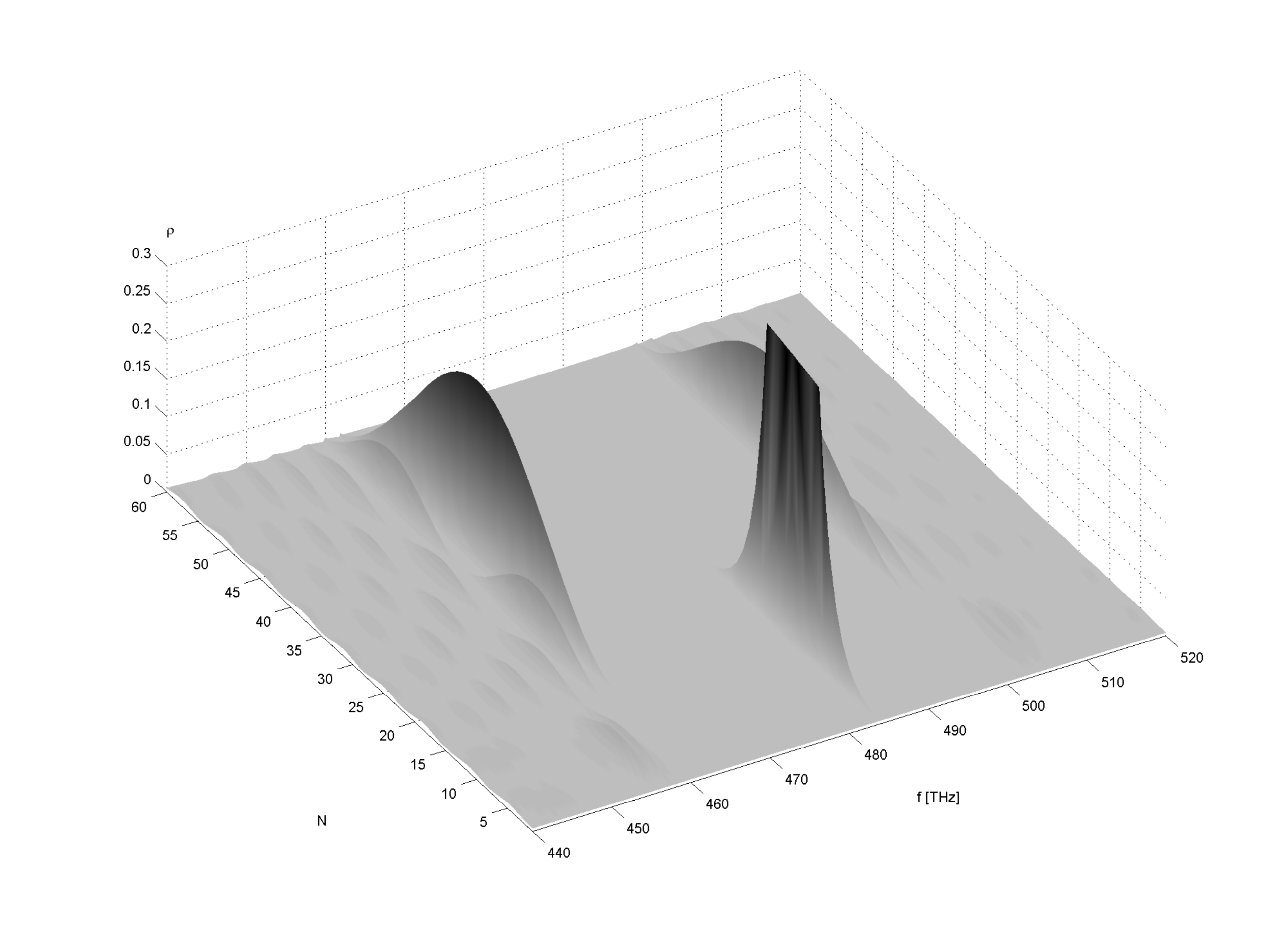}
\caption{Mode spectrum for perpendicular incidence in a PC with 1 defect versus frequency $f$ and number of elementary cell $N$.}\label{fdef1}
\end{center}
\end{figure}

Introducing a defect of width to a layer results in an appearance
of a~defect mode. The size of the defect determines at which
frequency does it happen -- see Fig. \ref{deffreq} -- obviously,
because a resonant frequency of a resonator depends on its width.
However, the defect mode appears only in a certain number of
layers near the defected one, because electromagnetic field with
frequency from a band gap has an envelope which vanishes in
consecutive layers -- it has a limited range of penetration, hence
the defect has no noticeable influence on mode structure of layer
which is sufficiently distant. This behaviour is depicted in Fig.
\ref{fdef1} (values on $z$-axis in this figure are limited).

\begin{figure}
\begin{center}
\includegraphics[width=.95\textwidth]{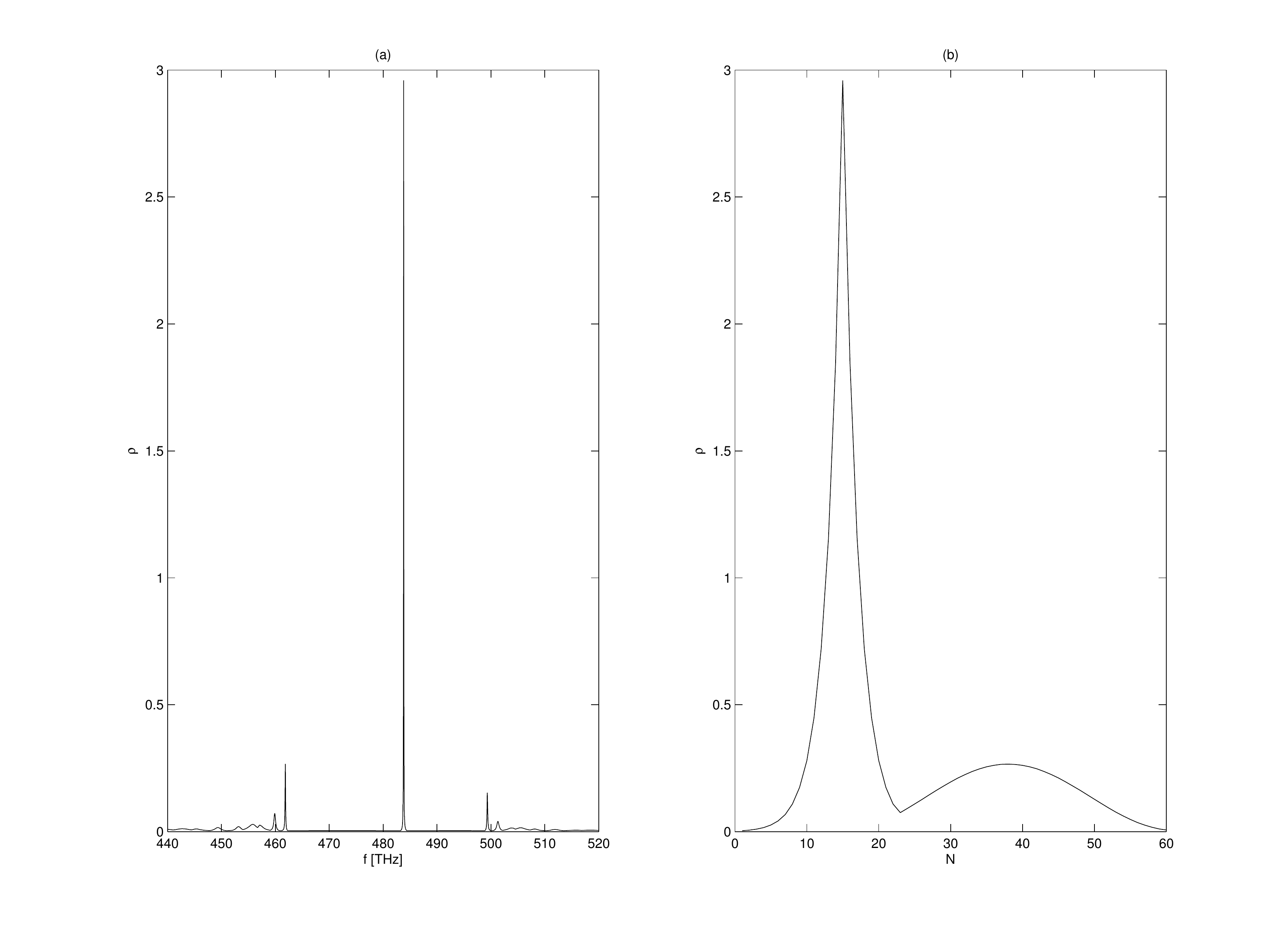}
\caption{Maximal values of mode spectrum in a PC with 1 defect versus (a) frequency $f$ and (b) number of elementary cell $N$ for defect in $15^\mathrm{th}$ elementary cell.}\label{fdef1max}
\end{center}
\end{figure}
\begin{figure}
\begin{center}
\includegraphics[width=.95\textwidth]{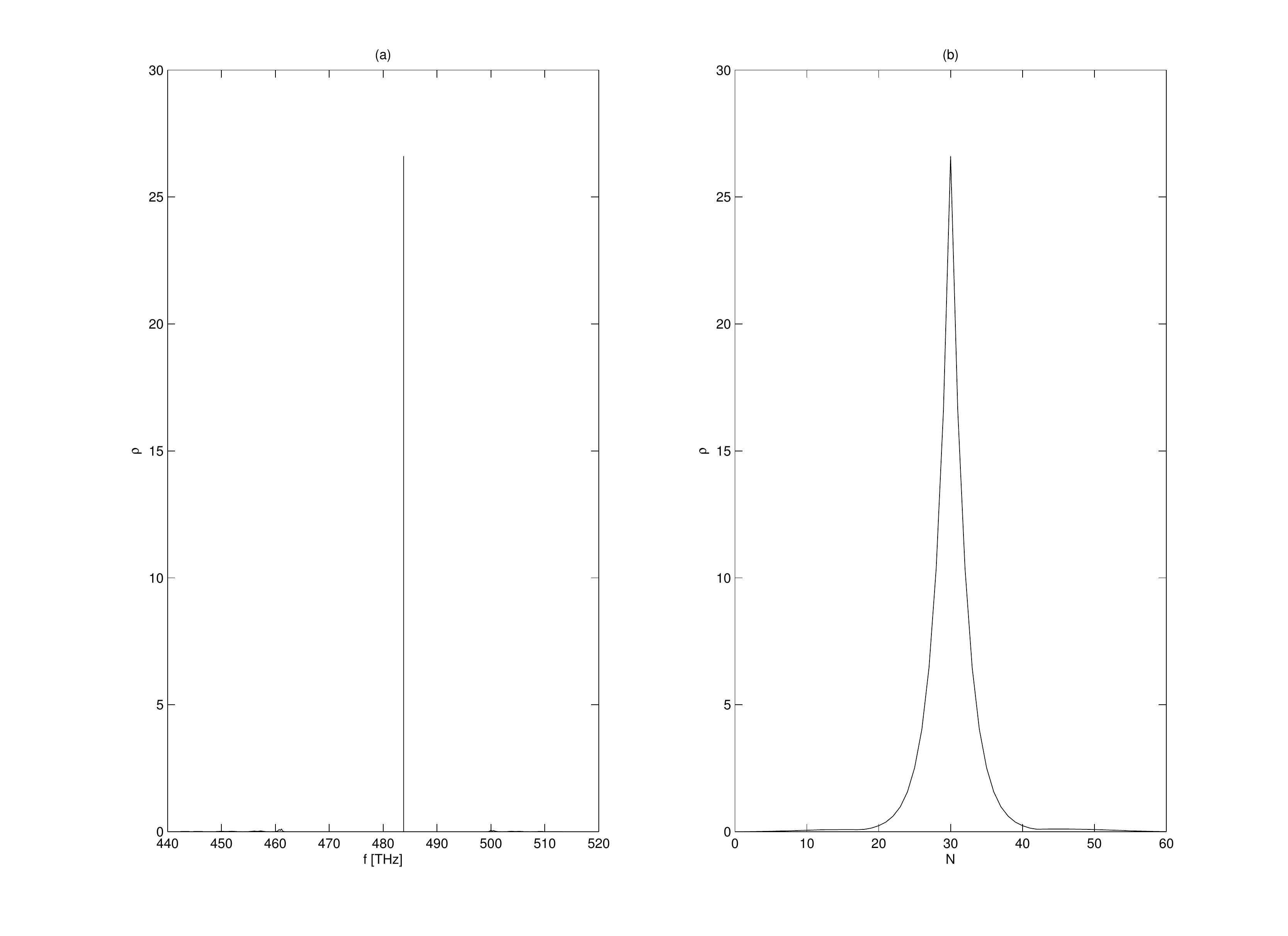}
\caption{Maximal values of mode spectrum in a PC with 1 defect versus (a) frequency $f$ and (b) number of elementary cell $N$ for defect in $30^\mathrm{th}$ elementary cell.}\label{fdef1maxcen}
\end{center}
\end{figure}

Figure \ref{fdef1max} shows maximal values of mode spectrum for
each frequency and in each layer. It can be seen, that the defect
introduces much higher values of mode spectrum than the edges of
the band gap. Modification of mode spectrum depends on which layer
is defected -- a similar graph for a defect in the middle of the
photonic crystal is shown in Fig. \ref{fdef1maxcen}. In this case
the defect mode frequency is the same, but values of mode spectrum
associated with the defect are even higher.

\begin{figure}
\begin{center}
\includegraphics[width=.95\textwidth]{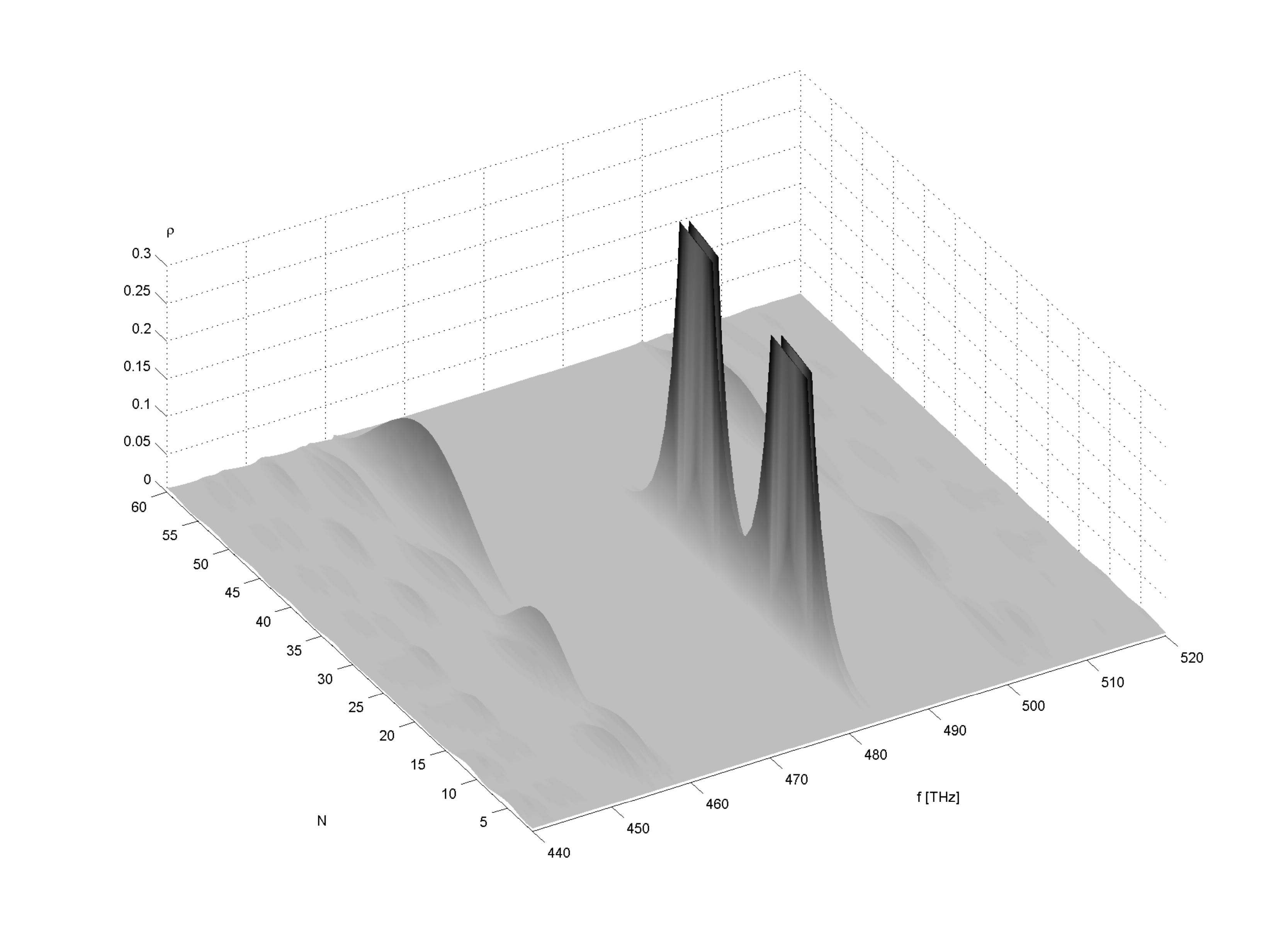}
\caption{Mode spectrum for perpendicular incidence in a PC with 2 defects versus frequency $f$ and number of elementary cell $N$.}\label{fdef2}
\end{center}
\end{figure}
\begin{figure}
\begin{center}
\includegraphics[width=.95\textwidth]{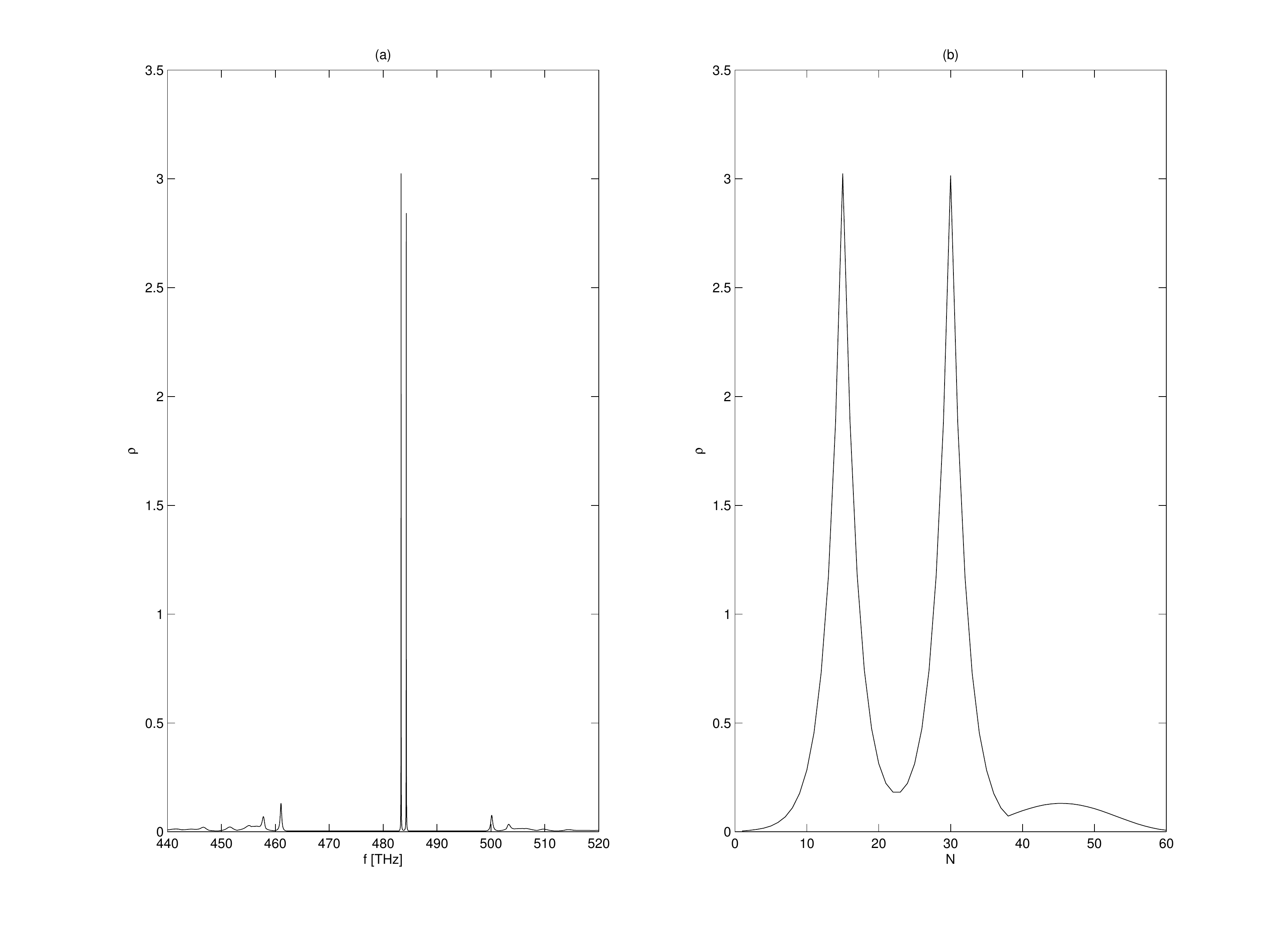}
\caption{Maximal values of mode spectrum in a PC with 2 defects versus (a) frequency $f$ and (b) number of elementary cell $N$.}\label{fdef2max}
\end{center}
\end{figure}
\begin{figure}
\begin{center}
\includegraphics[width=.95\textwidth]{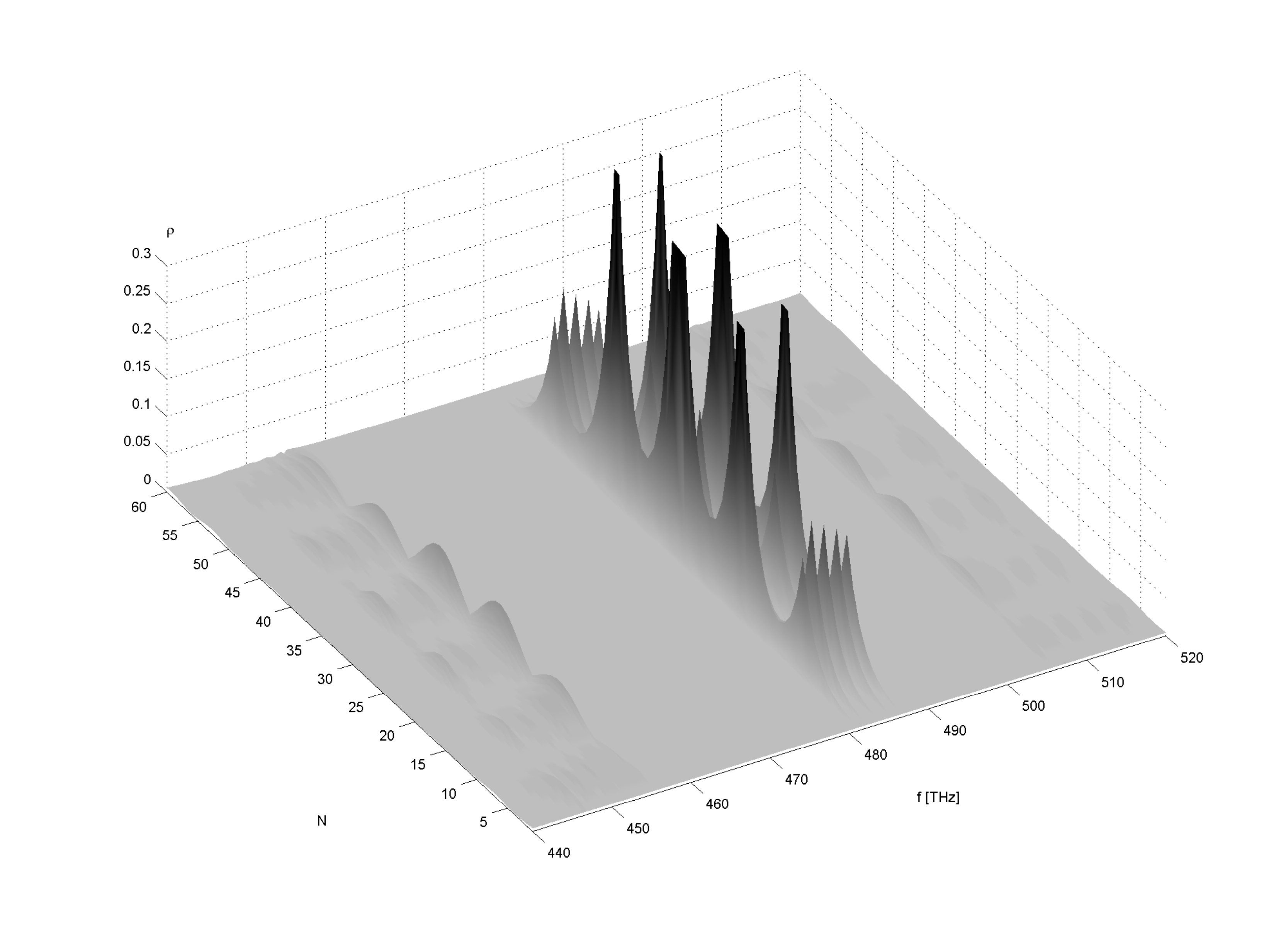}
\caption{Mode spectrum for perpendicular incidence in a PC with 5 defects versus frequency $f$ and number of elementary cell $N$.}\label{fdef5}
\end{center}
\end{figure}
\begin{figure}
\begin{center}
\includegraphics[width=.95\textwidth]{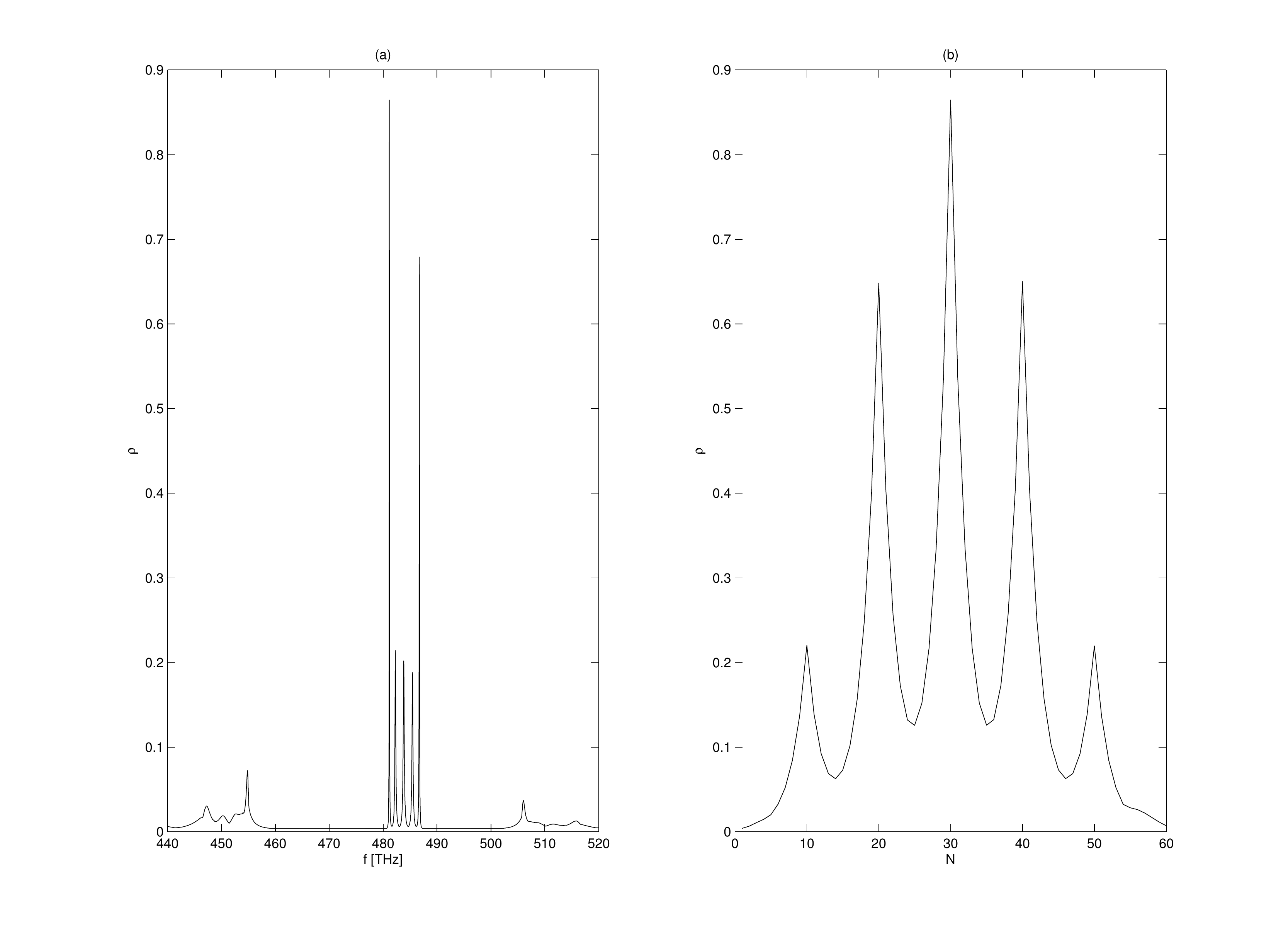}
\caption{Maximal values of mode spectrum in a PC with 5 defects versus (a) frequency $f$ and (b) number of elementary cell $N$.}\label{fdef5max}
\end{center}
\end{figure}

If a photonic crystal has two defected layers, then there appear
two regions in which influence of defect modes is significant.
Overlap of these regions results in two defect modes, with two
different frequencies, instead of one. This case is shown in Fig.
\ref{fdef2} and maximal values of mode spectrum are plotted in
Fig. \ref{fdef2max}. Adding more defect leads to further splitting
of the band gap by a new defect band -- see Figs. \ref{fdef5} and
\ref{fdef5max}.

\section{Conclusions}\label{Secconc}
In this paper we have analyzed the influence of defects in
a 1D photonic crystal on the density of states and
spontaneous emission, in both spacial and frequency domains. We
have used the effective resonator model and derived an analytical
expression for mode spectrum. With the help of this relation the
presented numerical results revealing properties of 1D photonic crystal with
defects have been obtained. In particular, it has been shown that
a single defect has local character and its influence on mode
structure diminishes with the distance from the defect position.

\begin{acknowledgements}
Project granted by the Ministry of Education and Science in
Poland, project no 3~T11B~069~28. The authors are members of the
Sixth European Union Framework Program for Research and
Technological Development (FP6) -- Network of Excellence on
Micro-Optics ``NEMO''.
\end{acknowledgements}

\end{article}
\end{document}